# Photoacoustic Tensile Imaging

Daohuai Jiang[1*], Xuanxuan Ye[1], Hengrong Lan[2,3], Xianzeng Zhang[1*], Fei Gao[2,3,4*]

1. Key Laboratory of Optoelectronic Science and Technology for Medicine, Ministry of Education, Fujian Provincial Key Laboratory for Photonics Technology, Fujian Normal University, Fuzhou, 350117, China
2. School of Biomedical Engineering, Division of Life Sciences and Medicine, University of Science and Technology of China, Hefei, Anhui, 230026, China
3. Hybrid Imaging System Laboratory, Suzhou Institute for Advanced Research, University of Science and Technology of China, Suzhou, Jiangsu, 215123, China
4. School of Engineering Science, University of Science and Technology of China, Hefei, Anhui, 230026, China

* Corresponding author.   jdaohuai@163.com; xzzhang@fjnu.edu.cn; fgao@ustc.edu.cn

# Abstract

Photoacoustic (PA) imaging combines the high optical absorption contrast of optical imaging with the deep tissue penetration of ultrasound detection, offering great potential for functional imaging and disease diagnosis. However, current PA imaging methods mainly explore optical absorption properties of biological tissue. To the best of our knowledge, tensile measurement based on PA effect is still an untapped area to be explored. In this work, we propose photoacoustic tensile imaging (PATI), a new PA imaging modality enabling quantitative assessment of tensile stress in biological samples. PATI exploits the nonlinear PA response induced by dual-pulse laser excitation to establish a mapping between the applied tension and the increment of the nonlinear PA signal. By varying the temporal delay between the heating and detecting laser pulses, the relationship between tensile force and nonlinear PA characteristics is quantitatively analyzed. Phantom experiments demonstrate a strong correlation between the nonlinear PA signal intensity and the applied tensile force. These results confirm the feasibility of the proposed approach for tensile force monitoring, which holds potential in biomedical applications, such as vascular pressure monitoring.

Keywords: nonlinear photoacoustic effect, photoacoustic tomography, dual-pulse excitation, tensile measurement

# Introduction

Photoacoustic imaging (PAI) combines the high optical absorption contrast with deep penetration capability of ultrasound detection, emerging as a powerful modality in biomedical imaging[1–3]. It is based on the mechanism of photoacoustic effect: pulsed laser irradiation induces localized thermoelastic expansion in biological tissues, generating broadband ultrasonic waves that can be detected and reconstructed to reveal spatially resolved optical absorption distributions[4]. This enables high-contrast, high-resolution imaging of structural, molecular, and functional information[5–7], with broad applications in tumor diagnosis[8–11], vascular assessment[12,13], and metabolic monitoring[14,15].

In recent years, the development of nonlinear PAI has further expanded the capability of PAI beyond its conventional linear regime[16]. Unlike traditional PA generation, which follows a linear relationship between absorbed optical energy and the initial pressure rise, nonlinear PA methods harness complex light–matter interactions, such as dual-pulse excitation, intensity modulation, and wavelength combination, which enhance imaging contrast, resolution, and functional specificity. For example, Grüneisen-relaxation photoacoustic microscopy introduced by Wang et al. exploits the temperature dependence of the Grüneisen parameter $\Gamma$ to achieve noninvasive temperature mapping[17]. Danielli et al. revealed nonlinear spectral responses of hemoglobin, enabling highly specific functional imaging[18]. Goy et al. further demonstrated resolution enhancement by modulating excitation intensity of laser[19], while Cho et al. advanced computational reconstruction through $p_{th}$ root spectral amplitude scaling to improve clinical PA/US image resolution[20]. More recently, Ke et al. achieved accurate quantification of high-speed blood flow using continuous-wave–assisted nonlinear PA techniques, highlighting the potential of nonlinear methods for microcirculation dynamics[21].

Dual-pulse nonlinear photoacoustic imaging (DP-NPAI) has become an important strategy for inducing and isolating nonlinear PA responses. By delivering two laser pulses within a time interval shorter than the tissue's thermal relaxation time, the first pulse elevates the local temperature and alters the Grüneisen coefficient, thereby amplifying the PA amplitude generated by the second pulse. The differential signal between the two pulses effectively extracts the nonlinear component, enabling suppression of background noise and enhancement of imaging contrast[22]. Tian et al. systematically demonstrated the feasibility of DP-NPAI in differentiating tissue compositions and applied it to lipid detection in liver phantoms[23]. Furthermore, DP-NPAI has been explored in molecular contrast enhancement [24] and hemodynamic monitoring[21], indicating its potential for high-specificity imaging in complex biological environments.

From a clinical perspective, mechanical stresses within intracranial vasculature serve as critical biomechanical indicators for a variety of neurological disorders. The risk of intracranial aneurysm rupture is strongly associated with aneurysmal wall stress distribution[25,26], and the development of hypertensive intracerebral hemorrhage is closely linked to the mechanical state of the vessel wall[27,28]. Existing clinical imaging modalities, such as ultrasound, CT, and MRI, lack the capability for noninvasive, real-time, and quantitative monitoring of intracranial vascular pressure or wall stress[25,29]. Although PA imaging have recently been applied to intracranial hemorrhage detection [30,31] and tumor-related vascular imaging[32], the relationship between vascular mechanical load and the nonlinear PA response has not yet been investigated. Therefore, there is an unmet clinical need for noninvasive tensile imaging of vasculature.

To fill this gap, we propose a new PA imaging modality, named photoacoustic tensile imaging (PATI), based on dual-pulse nonlinear PA measurement for quantitative characterization of tensile loading in biological samples.

## Methods

PATI regulates the nonlinear PA response by controlling the temporal delay Δt between two laser pulses. (as shown in Fig.1(b)-(c)) When the first pulse, which is defined as the heating pulse, irradiates the sample, it induces a transient temperature rise that does not completely dissipate before the arrival of the second pulse. The detecting pulse then interacts with this elevated temperature field, enhancing the thermoelastic expansion and generating a nonlinear PA signal. To ensure a significant nonlinear effect, the delay between the heating and detecting pulses must remain shorter than the thermal relaxation time of the absorber ($\Delta t \leq \tau_{th}$), satisfying the thermal confinement condition[22]. Within this interval, the residual heat deposited by the heating pulse modifies the local Grüneisen parameter $\Gamma$, thereby amplifying the PA amplitude generated by the detecting pulse. This temperature-dependent modulation of $\Gamma$ is known as the Grüneisen relaxation (GR) effect[17]. Therefore, dual-pulse excitation enables modulation of the Grüneisen parameter, providing additional contrast mechanisms for functional PA imaging. The schematic of the proposed PATI approach is shown in Fig. 1, which investigates the relationship between nonlinear PA signals under dual-pulse excitation and the applied mechanical load. This method offers a new strategy for imaging the mechanical properties of biological tissues.

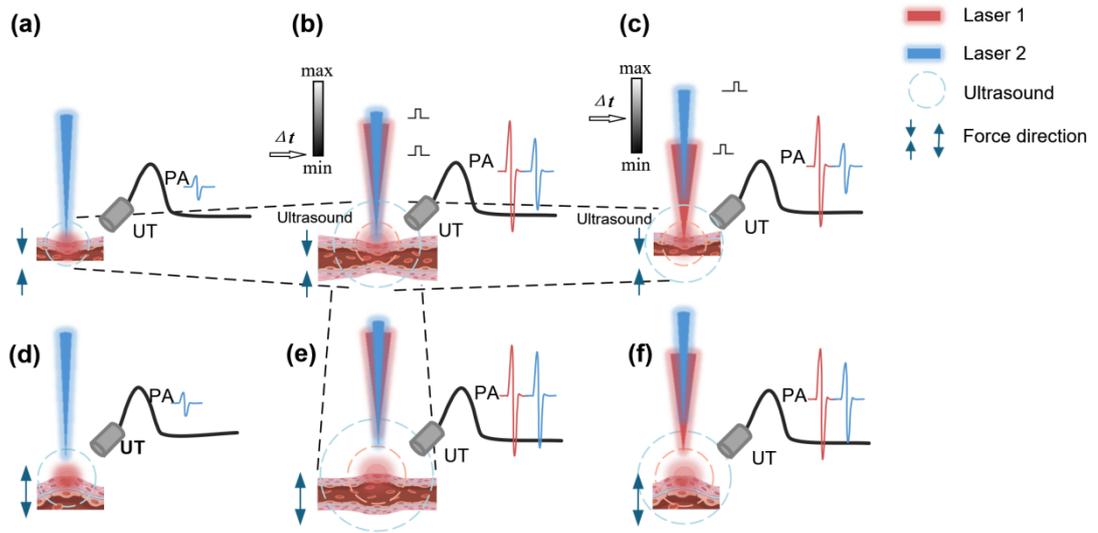

Fig. 1. Schematic illustration of dual-pulse PA tension imaging. (a) Conventional PA signal acquisition under single-pulse excitation; (b)-(c) PA signals generated by dual-pulse excitation with different inter-pulse delays; (d)-(f) Corresponding PA signal measurements in (a)-(c) under different loading directions applied to the phantom. PA: photoacoustic; UT: ultrasound transducer.

Under conventional single-pulse excitation, the initial PA pressure can be expressed as[33]:

$$p_0 = \Gamma \eta_{th} \mu_a F \qquad (1)$$

where $\eta_{th}$ is the heat conversion efficiency, $\mu_a$ is the optical absorption coefficient, $F$ is the optical fluence. The Grüneisen parameter $\Gamma$, a dimensionless quantity reflecting the thermoelastic properties of tissue, is defined as:

$$\Gamma = \frac{\beta v_s^2}{C_P} \qquad (2)$$

where $\beta$ is the thermal expansion coefficient, $v_s$ is the speed of sound, and $C_p$ is the specific heat capacity at constant pressure. Among these parameters, $\beta$ exhibits the strongest dependence on temperature. Therefore, when the local temperature increases from the baseline $T_0$ by $\Delta T$, $\Gamma$ increases approximately linearly[17]:

$$\Gamma(T_0 + \Delta T) \approx \Gamma_0 + \Gamma' \cdot \Delta T \qquad (3)$$

where $\Gamma_0$ is the baseline Grüneisen parameter and $\Gamma'$ denotes its temperature derivative.

With dual-pulse excitation, the heating pulse produces an initial temperature rise satisfying $\Delta T \propto \mu_a F_1$, where $F_1$ is the fluence of the heating pulse. Because heat diffusion requires finite time, the temperature decays gradually after the heating pulse. According to the model of Tian et al.[22], the change in the Grüneisen parameter after a delay $\Delta t$ is: $\Delta \Gamma(\Delta t) = \Gamma' \cdot \Delta T \cdot \tau(\Delta t)$, where $\tau(\Delta t)$ is the thermal relaxation function, and $\Delta T \cdot \tau(\Delta t)$ represents the remaining temperature rise at time . Thus, the effective Grüneisen parameter becomes: $\Gamma(\Delta t) = \Gamma_0 + \Delta \Gamma(\Delta t)$. The detecting pulse therefore produces an amplified PA signal $V_2'$ compared with the unheated condition $V_2$. The nonlinear PA increment is given by [22,34] (as shown in Fig 1):

$$\Delta V_2(\Delta t) = V_2' - V_2 = \Delta \Gamma(\Delta t) \eta_{th} \mu_a F_2 \tag{4}$$

where $F_2$ is the fluence of detecting pulse.

Since $\Delta \Gamma(\Delta t) \propto \Delta T \propto F_1$, the increment follows $\Delta V_2(\Delta t) \propto F_1 F_2$, and its amplitude decay curve over $\Delta t$ directly reflects the thermal relaxation dynamics of the absorber. This relaxation behavior is governed by the thermal diffusivity $\alpha = \frac{k}{\rho C_p}$, determined by thermal conductivity $k$, density $\rho$, and heat capacity $C_p$. Meanwhile, acoustic parameters such as $v_s$ influence $\Gamma$ as described by Eq. (2). Thus, externally applied tension $f$ alters both the thermal and acoustic behavior of the sample, modifying the nonlinear PA amplitude and relaxation kinetics.

To establish the relationship between the applied tensile force $f$ and the nonlinear PA signal, we further analyzed the stress-induced variations in material parameters within the imaged sample. When a uniaxial tensile force $f$ is applied, the internal stress can be expressed as: σ=$f/A$, where $A$ denotes the cross-sectional area of the sample. Within the elastic regime and under small-strain conditions, the acoustic and thermal parameters of the medium can be approximated as linear functions of stress [35–38]:

$$v_s^2(σ) \approx v_{s0}^2(1+ξσ) \tag{5}$$

$$k(σ) \approx k_0(1+κσ) \tag{6}$$

where $v_{s0}$ and $k_0$ represent the sound speed and thermal conductivity at zero stress, respectively, and ξ and κ are material-dependent stress coefficients. Assuming that the thermal expansion coefficient $β$ and the specific heat capacity are weakly dependent on stress, substituting Eq. (5) into Eq. (2) yields the stress dependence of the Grüneisen parameter:

$$\Gamma(σ) = \frac{β[v_s(σ)]^2}{c_P} \approx \Gamma_{σ0}(1 + ξσ) \tag{7}$$

where $\Gamma_{σ0}$ denotes the baseline Grüneisen parameter under zero stress, i.e., $\Gamma_{σ0}=\Gamma_0=\Gamma(σ=0, T=T_0)$, Accordingly, the temperature derivative of the Grüneisen parameter also varies with stress:

$$\Gamma'(σ) = \frac{\partial \Gamma(σ)}{\partial T} \approx \Gamma_0'(1 + ξσ) \tag{8}$$

where $\Gamma_0' = \frac{\partial \Gamma_0}{\partial T}$. Similarly, the thermal diffusivity is modulated by stress:

$$α(σ) \approx α_{σ0}(1+κσ) \tag{9}$$

where $\alpha_{\sigma 0}=\frac{k_0}{\rho C_p}$ is the thermal diffusivity under zero stress.

As a result, in the dual-pulse PA model, both the amplitude and relaxation rate of the nonlinear PA signal increment $\Delta V_2$ are stress dependent. Specifically, the PA tensile sensitivity coefficient $V_{ts}$ is proportional to $\Gamma'$, while the relaxation rate parameter $b$ is proportional to $\alpha$. Their relationships with stress $\sigma$ and tensile force $f$ can thus be expressed as:

$$V_{ts}(\sigma) \propto (1+\xi\sigma)=1+\xi\ (f\,/\,A) \qquad (10)$$

$$b(\sigma) \propto (1+\kappa\sigma)=1+\kappa\ (f\,/\,A) \qquad (11)$$

To further relate the decay behavior of $\Delta V_2(\Delta t)$ to the tensile-dependent parameters $V_{ts}$ and $b$, the nonlinear PA signal increment was modeled as:

$$\Delta V_2(\Delta t) = V_{of} + V_{ts}(\gamma \Delta t - 1)e^{-b\Delta t} \qquad (12)$$

where $e^{-b\Delta t}$ represents the thermal-diffusion-driven decay; ($\gamma \Delta t - 1$) characterizes initial transient behavior, where $\gamma$ is the generation rate constant; $V_{of}$ is a delay-independent offset; The parameters $V_{ts}$ and $b$ are the key indicators of mechanical state, quantifying the nonlinear PA signal amplitude and thermal relaxation rate, respectively. Among these parameters, $V_{ts}$, which directly originates from the signal amplitude, is easier to measure and exhibits a stronger and more explicit linear dependence on stress. Therefore, it is selected as the primary metric for tensile force characterization. Based on Eq. (10), within the small-strain regime, the fitted $V_{ts}$ satisfies a linear relationship with the applied tensile force:

$$V_{ts}(f)=V_{ts0}+S\cdot f \qquad (13)$$

where $V_{ts0}$ is the value of $V_{ts}$ at zero tensile force, and $S$ is the correction coefficient for $V_{ts}$ and tensile force. Consequently, by controlling the inter-pulse delay $\Delta t$, measuring the nonlinear PA signal increment $\Delta V_2(\Delta t)$, and performing curve fitting and calibration using Eqs. (12)-(13), the tensile-dependent characteristic parameter $V_{ts}(f)$ can be extracted, enabling quantitative discrimination and assessment of the mechanical state of the sample.

## Experimental

The experimental setup is shown in Fig. 2. The system employed two Q-switched pulsed lasers operating at 532 nm with pulse durations of approximately 8-9 ns and a repetition rate of 10 Hz. Laser-1 served as the heating source, while Laser-2 was used to generate the nonlinear PA signal. The two laser beams were spatially combined and directed onto the sample under investigation, as illustrated in Fig. 2(a).

A mechanical tensile loading device coupled with a force gauge was used to apply and monitor controlled tensile forces on the sample. For single channel PA signal detection, a single-element ultrasound transducer (center frequency: 5 MHz, bandwidth: 60%) was used to detect PA signals, which were amplified by a low-noise amplifier and recorded using a digital oscilloscope with a sampling rate of 8 GSa/s and a bandwidth of 200 MHz. For photoacoustic compute tomography (PACT) imaging, the detection system was replaced with a 128-element linear array transducer (center frequency: 7.5 MHz) and a data acquisition system (DAQ) operating at a sampling rate of 40 MSa/s, as shown in Fig. 2(b). All other system configurations remained unchanged.

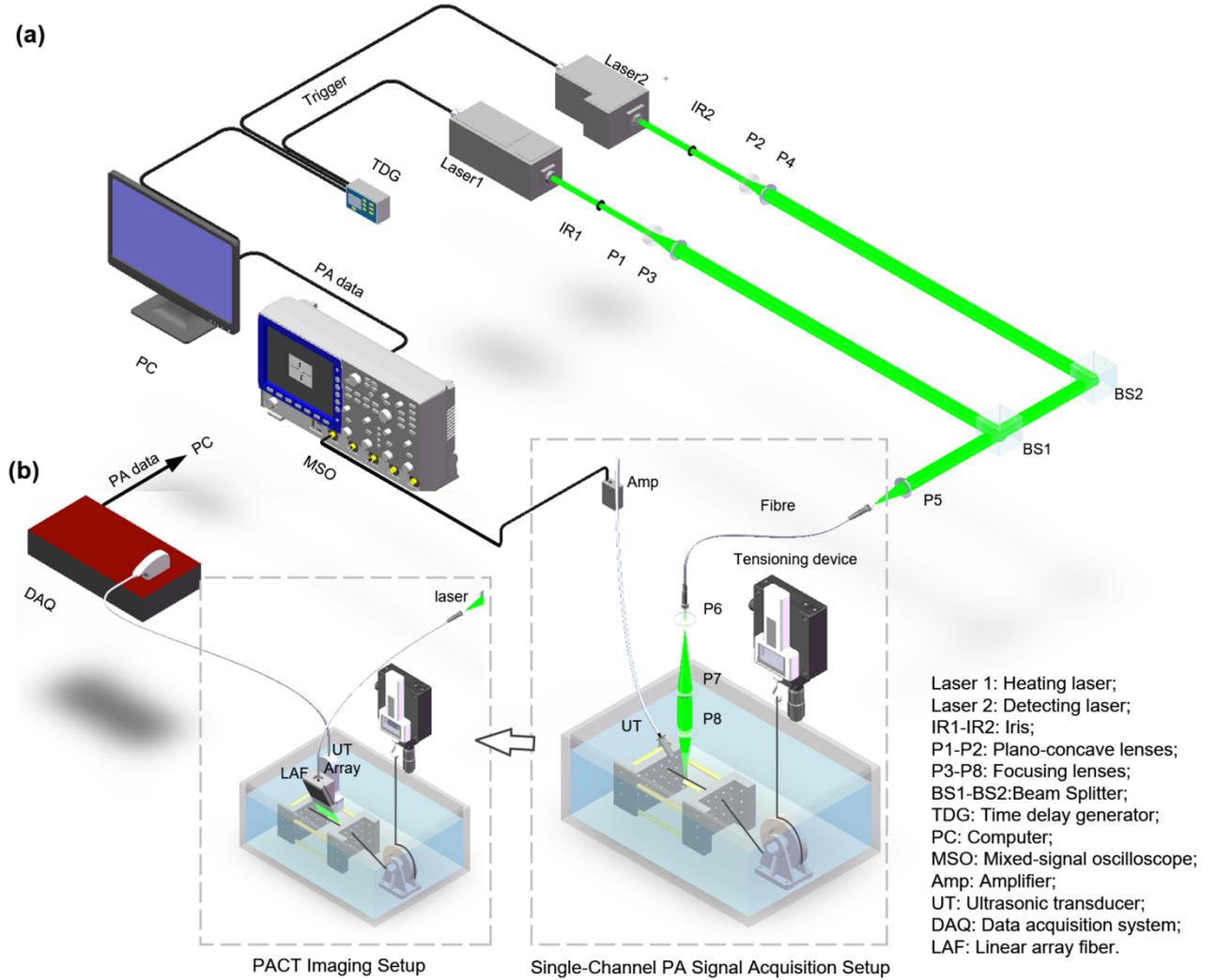

Fig. 2. The system setup of PATI for tensile measurement. (a) the setup of dual-pulse nonlinear PA signal acquisition with single channel. (b) the detection and sampling setup of PACT.

To precisely control the temporal delay Δt between the two laser pulses, a four-channel programmable delay generator was employed. The delay generator controls the timing of the heating and detecting laser pulses, which sequentially excite the sample to generate PA signals. The signals are amplified using low-noise analog conditioning and then digitized. In the digital domain, time-alignment correction is applied to compensate for propagation and electronic delays. The recorded PA waveform is separated into heating and detecting pulse induced components, and the detecting-pulse amplitude is extracted for calculating the nonlinear PA increment and analysis.

A tungsten wire with a diameter of 100 μm was used as the tensile test sample, with the applied force set and measured using the tensile loading apparatus. Both nonlinear PA signal detection and PACT imaging experiments were conducted. In the single-point detection experiments, PA responses of the tungsten wire were measured under different tensile forces (0.29 N, 1.26 N, 2.31 N, and 3.83

N). The pulse energies delivered to the sample were 2.23 mJ for the heating pulse and 0.86 mJ for the detecting pulse, with a focal spot diameter of approximately 3 mm. The dual-pulse delay Δt was varied from 0 to 200 ns with a step size of 2 ns.

For PACT imaging, the fiber output was configured as a linear illumination source, with pulse energies of approximately 45 mJ and 17 mJ for the heating and detecting pulses, respectively. Static imaging experiments were performed under tensile forces of 0 N, 2.5 N, and 3.9 N, with Δt set to 0, 15, 100, and 200 ns. For each condition, 50 consecutive frames were acquired, and stable signals were reconstructed using the delay-and-sum (DAS) algorithm. In dynamic imaging experiments, the tensile force applied to the tungsten wire was continuously varied while Δt was fixed at 100 ns, enabling real-time PACT imaging to evaluate the relationship between applied tension and PATI images.

## Results

By varying the inter-pulse delay Δt, detecting-pulse induced PA signals under different preheating conditions were acquired. Representative raw waveforms (Fig. 3(a)) show that the heating pulse significantly enhances PA amplitude. As Δt increases, the enhancement gradually weakens, and the PA amplitude decreases, consistent with the theoretical behavior predicted by Eq. (4). When Δt exceeds the thermal confinement condition, the nonlinear interaction diminishes, and the two pulses generate independent PA signals.

To enable accurate quantitative analysis, the measured PA amplitudes were normalized by laser fluence and corrected for low-frequency baseline drift. The resulting nonlinear PA increments as a function of Δt under different tensile forces are shown in Fig. 3(b). The nonlinear PA increment exhibits a pronounced delay-dependent behavior, with higher tensile forces producing consistently larger increments within 12 ns < Δt ≤ 200 ns. In addition, the inflection points of the delay-dependent curve shifts toward shorter Δt as the applied tensile force increases, indicating a force-dependent modulation of the nonlinear PA response.

The fitted parameters extracted from Eq. (5) were further analyzed with respect to tensile force (Fig. 3(c), Table 1). Among all parameters, the nonlinear PA intensity coefficient $V_{ts}$ shows the strongest linear correlation with tensile force, with a coefficient of determination $R^2$=0.9913. The thermal relaxation rate constant $b$ exhibits a moderate correlation ($R^2$=0.8127), whereas the offset term $V_{of}$ and generation rate constant $\gamma$ show weak sensitivity to tensile force. These results confirm that the nonlinear PA signal with dual-pulse excitation provides a robust metric for tensile force quantification.

The imaging performance of PATI was first evaluated under identical tensile conditions (0 N). As shown in Fig. 4(a), nonlinear PA images reconstructed at Δt = 0 ns exhibit significantly higher contrast than conventional linear images, with enhanced signal intensity at representative pixel locations. The effect of inter-pulse delay on nonlinear PA imaging was then investigated. Incremental nonlinear PA images acquired at different delays under *0 N* tensile force (Fig. 4(e)) show a gradual reduction in image contrast with increasing Δt, consistent with the weakening nonlinear PA effect. At fixed delays of 100 ns and 200 ns, increasing tensile force leads to improved target visibility and enhanced image contrast (Fig. 4(f)-(g)), accompanied by a corresponding increase in nonlinear PA increments (Fig. 4(i)-(j)). Therefore, the tensile state of the investigated sample can be directly inferred through image analysis of PATI.

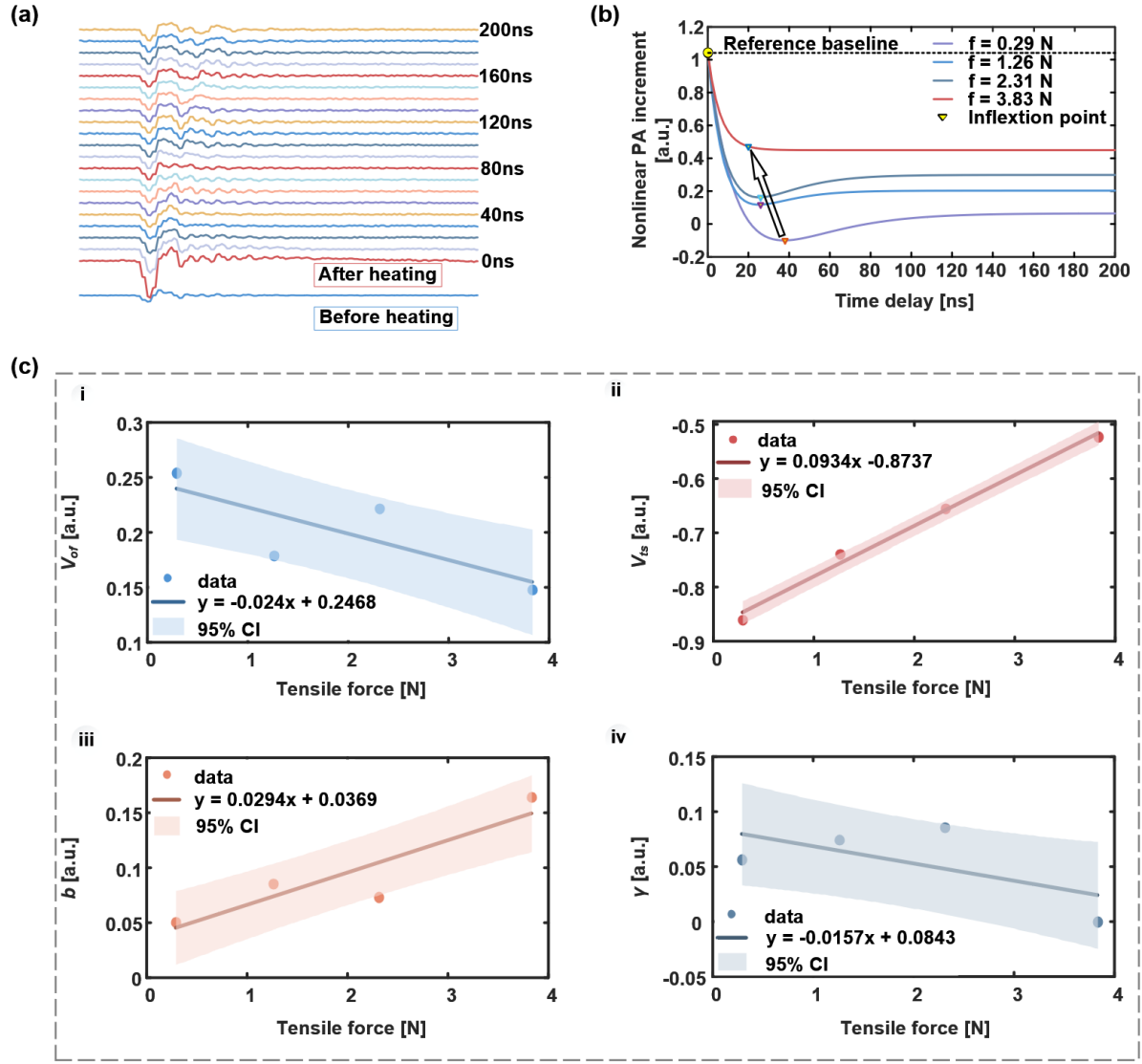

Fig. 3. Dual-pulsed induced PA signals and correlation analysis. (a) The PA signals generated by dual pulses with varying inter-pulse delays.(b) Fitted curves of the nonlinear PA increment as a function of the inter-pulse delay with different tensile forces.(c) The functional relationship between the key parameters and the tensile force f in the empirical model constructed.

**Tab.1 Goodness-of-fit statistics for the relationship between key parameters and tensile force $f$**

| parameters | correlation coefficient (r) | Significance (p) | coefficient of determination ($R^2$) |
| --- | --- | --- | --- |
| $V_{of}$ | -0.7772 | 0.2228 | 0.6040 |
| $V_{ts}$ | 0.9956 | 0.0044 | 0.9913 |
| $b$ | 0.9015 | 0.0985 | 0.8127 |
| $\gamma$ | -0.6250 | 0.375 | 0.3907 |

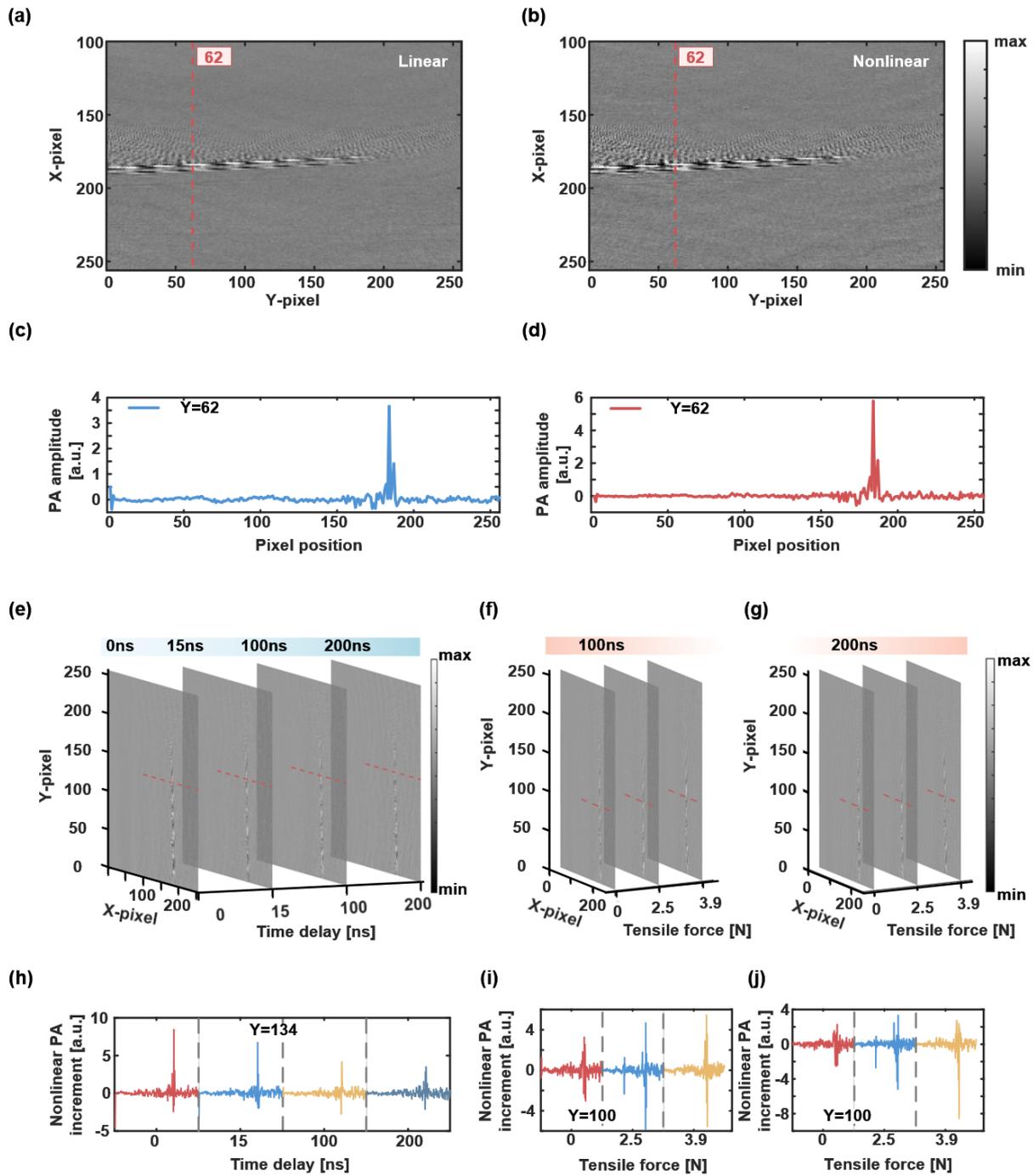

Fig.4. The comparison of PATI results. (a) Conventional PACT image; (b) Dual-pulse nonlinear PA image; (c) and(d) PA images pixel value at position Y=62 corresponding to (a) and (b). (e) Incremental nonlinearity PA image comparison at different delay times (0 ns, 15 ns, 100 ns, 200 ns) under the same tensile force condition (0 N); (f)-(g) PATI results at different tensile forces with delay times of 100 ns and 200 ns. (h) Nonlinear PA increment for the delay times corresponding to (e). (i)-(h) Nonlinear PA increment for the delay times corresponding to (f)-(g).

To further examine the dynamic response of PATI to tensile loading, time-resolved measurements were performed at a fixed delay time of 100 ns under continuously varying tensile force. As shown in Fig. 5(a), the applied tensile force and the PA amplitude of PATI exhibit synchronized non-monotonic behavior, with an initial increase followed by a decrease.

A representative PA image acquired at t = 18.6 s and the corresponding PA signal from the pixel column with optimal contrast (Y=54) are shown in Fig. 5(b) and Fig. 5(c). The distribution of optimal-contrast pixel positions (Fig. 5(d)) is concentrated within Y = 52-56, indicating good spatial consistency. As shown in Fig. 5(e), the positive PA pulse peak closely follows the variation in peak-to-peak amplitude, whereas the negative peak remains relatively stable, indicating that the PATI's PA amplitude response is dominated by the positive pulse component. The scatter plot and trend line in Fig. 5(f)-(g) further confirm a strong positive correlation between tensile force and PATI's PA amplitude.

Detailed results of dual-pulsed induced dynamic PATI of tensile measurements are available in the Supplementary Information.

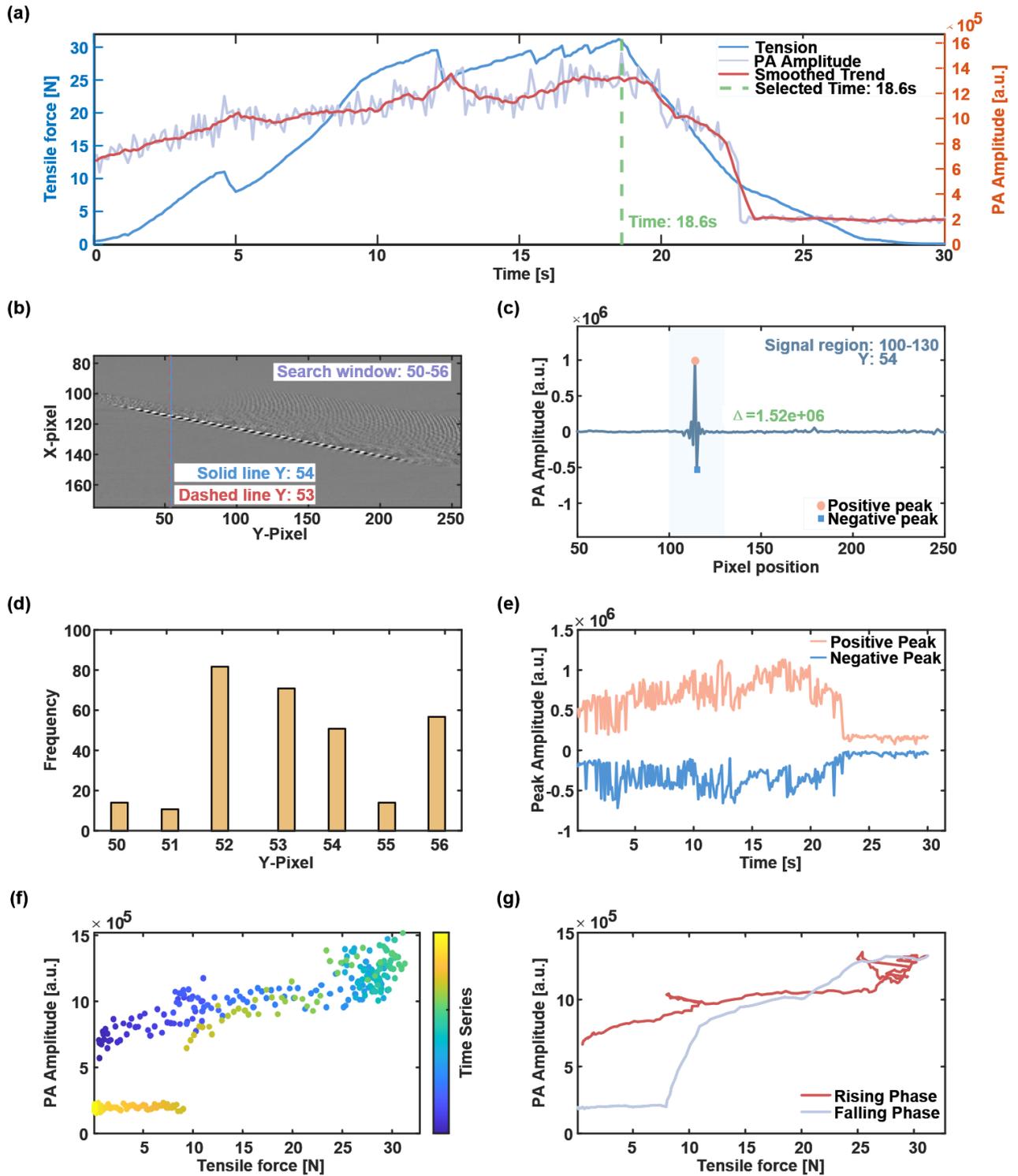

Fig. 5. Dynamic response characteristics of PATI's PA signal under continuous tensile stress. (a) Time-dependent curves of tensile force versus PATI's PA amplitude; (b) PATI image at 18.6 s. The red dashed line indicates the reference position, while the blue solid line marks the optimal contrast position. The search window covers ±3 pixels around the reference position. (c) PATI's PA signal from the pixel column with optimal contrast as marked in (b); (d) Frequency distribution of Y-pixel positions exhibiting optimal contrast; (e) Time-dependent curves of positive and negative PA peak amplitudes; (f) Relationship between PATI's PA amplitude and tensile force (scatter plot); (g) Relationship between PATI's PA amplitude and tensile force (smoothed data).

# Discussion

In this study, the PATI was used to investigate the relationship between tensile loading and thermal relaxation behavior of the medium. Experimental results obtained from tungsten wire samples demonstrate that the nonlinear PA increment exhibits a clear decay pattern with inter-pulse delay time, and within a specific delay range, the PA amplitude shows a strong correlation with the applied tensile force.

Although the absolute decay curves vary across samples, their overall trends remain consistent. Notably, when the applied tensile force drives the sample beyond the elastic deformation regime, the decay behavior changes systematically, indicating that the proposed method is sensitive to variations in mechanical state and capable of detecting mechanical transitions.

The fitting of the decay curves further reveals that, within the elastic range, the nonlinear intensity coefficient $V_{ts}$ exhibits a strong linear correlation with tensile force. This observation is consistent with the underlying mechanism that mechanical loading alters the microstructure of the medium and modulates the temperature dependence of the Grüneisen parameter. In addition, the thermal relaxation rate constant $b$ tends to increase with tensile force, suggesting accelerated heat diffusion under mechanical stretching. In contrast, the offset term $V_{of}$ and the generation rate constant $\gamma$ show limited sensitivity to tensile loading, likely reflecting background or secondary effects.

From an imaging perspective, nonlinear PA imaging based on the proposed dual-pulse excitation effectively translates tensile-induced signal variations into image contrast changes, enabling spatially resolved assessment of mechanical loading. These results demonstrate the feasibility of using nonlinear PA imaging for tensile evaluation and provide a methodological basis for PACT-based mechanical or pressure imaging.

# Conclusion

This study proposes a PATI method for dynamic tensile force measurement. By precisely controlling the inter-pulse delay between two laser excitations, the thermal relaxation dynamics of PA signals under different tensile states were characterized, and a model linking tensile force to nonlinear PA increment and delay time was established. This framework enables sensitive and quantitative characterization of mechanical loading.

The proposed method not only quantifies continuous variations in mechanical state but also effectively identifies the transition from elastic to plastic deformation through systematic changes in nonlinear signal behavior. Moreover, PATI provides enhanced contrast and signal-to-noise ratio, while visually revealing tensile-induced structural changes, thereby demonstrating the intrinsic coupling between nonlinear photoacoustic responses and mechanical properties.

The feasibility of the method was validated in phantom experiments, laying an important foundation for the development of noninvasive blood pressure and intracranial pressure imaging techniques. The underlying principle also holds promise for extension to other biomedical applications requiring high-precision mechanical characterization.